\documentclass[aps,prl,reprint,groupedaddress]{revtex4-1}

\usepackage{graphicx}  % needed for figures
\usepackage{dcolumn}   % needed for some tables
\usepackage{bm}        % for math
\usepackage{amssymb}   % for math

\begin{document}

\title{Post-transcriptional regulation of noise in protein distributions
during gene expression}
\author{Tao Jia}
\email{tjia@vt.edu}
\author{Rahul V. Kulkarni}
\email{kulkarni@vt.edu} 
\affiliation{Department of Physics, \\
Virginia Polytechnic Institute and State University, \\
Blacksburg, VA 24061}
\date{\today}

\begin{abstract}
The intrinsic stochasticity of gene expression can lead to large
variability of protein levels across a population of
cells. Variability (or noise) in protein distributions can be
modulated by cellular mechanisms of gene regulation; in particular,
there is considerable interest in understanding the role of
post-transcriptional regulation. To address this issue, we propose and
analyze a stochastic model for post-transcriptional regulation of gene
expression. The analytical solution of the model provides insight into
the effects of different mechanisms of post-transcriptional regulation
on the noise in protein distributions.  The results obtained also
demonstrate how different sources of intrinsic noise in gene
expression can be discriminated based on observations of regulated
protein distributions.
\end{abstract}

\maketitle 

%%%%%%%%%%%%%%%%%%%%%%%%%%%%%%%%%%%%%%%%%%%%%%%%%%%%%%

The intrinsic stochasticity of biochemical reactions involved in gene
expression can lead to large variability of protein levels across a
clonal population of cells \cite{kaern05}.  The need to regulate this
variability (or noise) in protein distributions places important
constraints on cellular pathways; in particular those that bring about
global changes in gene expression. In such pathways, recent research
has increasingly highlighted the role of {\it post-transcriptional}
control by proteins or regulatory small RNAs\cite{waters09}. An
important question that arises is: what is the role of
post-transcriptional regulation in controlling the noise in protein
distributions? This work addresses the preceding question in the
context of stochastic models of gene expression as discussed below.

\par

To elucidate the source of intrinsic noise in protein distributions,
two coarse-grained stochastic models (Fig.1 A,B) have been proposed
\cite{kaufmann07}.  In one case, mRNA synthesis is modeled as a
Poisson process (Poisson scenario) and high variability in protein
levels is related to low abundance and infrequent synthesis of mRNAs
\cite{thattai01}. In the other case (Telegraph scenario), the promoter
is assumed to switch between active and inactive states, with mRNA
synthesis occuring in bursts only when the promoter is active
\cite{raser04,karmakar04,paulsson05,iyer09}.  The Poisson and
Telegraph scenario models both explain experimental observations of
noise in protein distributions \cite{bareven06,newman06,yu06} in terms
of 'bursts' of protein expression; however they make distinct
predictions for the underlying mRNA burst distribution (Fig.1). For
the Poisson scenario model, each burst of protein expression arises
from translation from a {\em single} mRNA; whereas for the Telegraph
scenario model, multiple mRNAs can contribute to a single burst of
protein expression. However, observations of protein burst
distributions alone do not uniquely identify the underlying mRNA burst
distribution and thus cannot be used to infer the source of intrinsic
noise in gene expression \cite{ingram08}. A currently open question is
whether this observation remains valid when protein burst
distributions are modulated by post-transcriptional regulation.

\par

\begin{figure}[tb]
%\begin{center}
\resizebox{7.5cm}{!}{\includegraphics{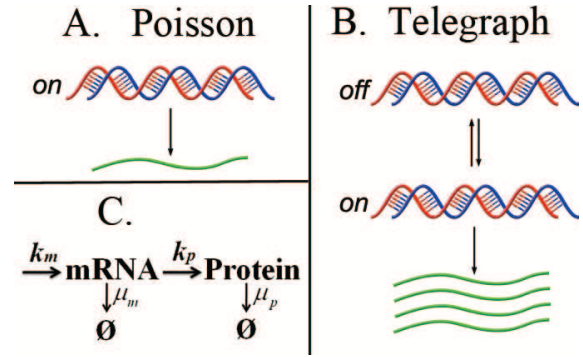}}
\caption{(A) Schematic representation of the Poisson scenario for mRNA
synthesis. Observed
bursts correspond to proteins produced from a single mRNA. (B) The
Telegraph scenario assumes that DNA transitions between an 'off' and
'on' state. mRNA synthesis only occurs in the 'on' state and mutiple
mRNAs can be produced per burst. (C) The Poisson scenario reaction scheme
for protein synthesis.\label{fig1}}
%\end{center}
\end{figure}\noindent

In this paper, we develop a framework for modeling
post-transcriptional regulation of genes expressed according to the
Poisson and Telegraph scenarios. The analytical results obtained
provide insight into how different mechanisms of post-transcriptional
regulation modulate noise in protein distributions. The results
obtained also show how different sources of intrinsic noise in gene
expression can be discriminated based on observations of {\em
regulated} protein burst distributions.

\par

We first analyze how post-transcriptional regulation modifies protein
bursts under the Poisson scenario.  Several recent studies have
focused on the corresponding mean-field and stochastic models, in
particular for regulation by small RNAs \cite{levine07, mitarai07,
mehta08}.  These studies have primarily focused on small RNAs which
act via irreversible stoichiometric degradation of mRNAs. On the other
hand, binding of the post-transcriptional regulator can more generally
be considered as a reversible reaction.  For this case, and in the
limit of large regulator concentrations (see below), we wish to
analyze the effects of different mechanisms of post-transcriptional
regulation on the noise in protein burst distributions.

\par

The proposed reaction scheme for our model is shown in Fig.(2A) for a
given concentration of the post-transcriptional regulator: the
regulator binds mRNA to form a complex with rate $\alpha$; the
dissociation rate for the complex is $\beta$.  The parameters $k_{p1}$
and $k_{p2}$ are the rates of protein production from the mRNA in free
and bound states and $\mu_{m}$ and $\mu_{c}$ are the corresponding
decay rates.  For global post-transcriptional regulators that are
present in large numbers, it is a good approximation that binding to
the target mRNA does not significantly alter the concentration of the
regulator. In this case, fluctuations in regulator concentration can
be neglected and the rate $\alpha$ can be taken to be constant.

\par

For the Poisson scenario, the protein burst distribution, $P_{b}(n)$,
corresponds to the number of proteins translated from a single mRNA
before it decays. During this process, the mRNA can exist in two
states: either free or bound in a complex with the
post-transcriptional regulator. Correspondingly, we define the
functions $f_1(n,t)$ and $f_2(n,t)$ (generalizing the approach
outlined in \cite{berg78}) which denote the probabilities of finding
the mRNA in free and bound states respectively at time $t$, having
produced a burst of $n$ proteins. The initial condition corresponds to
creation of the mRNA in its free state at $t=0$. Now, the burst
distribution $P_b(n)$ can be obtained from $f_1(n,t)$ and $f_2(n,t)$
as
\begin {equation}
P_b(n)=\int_{0}^{\infty}{f_1(n,t)}\mu_m\,dt +
\int_{0}^{\infty}{f_2(n,t)}\mu_c\,dt\label{eq:Pn_fr}
\end {equation}
Furthermore, the time evolution of $f_1(n,t)$ and $f_2(n,t)$ is
determined by the following Master equations:
\begin {eqnarray}
\frac{\partial f_1(n,t)}{\partial t} &=& k_{p1}(f_1(n-1,t)-f_1(n,t)) \nonumber \\
&-&(\mu_m+\alpha)f_1(n,t)+\beta f_2(n,t) \nonumber \\
\frac{\partial f_2(n,t)}{\partial t} &=& k_{p2}(f_2(n-1,t)-f_2(n,t)) \nonumber \\
&-&(\mu_c+\beta)f_2(n,t)+\alpha f_1(n,t) \label{eq:fr_2} 
\end {eqnarray}

\par

The above equations can be analyzed further using a combination of generating
functions and Laplace transforms. Specifically, defining $G_b(z) =
\sum_{n}{z^nP_b(n)}$ and
$F_{1,2}(z,s)=\sum_{n}{z^n\int_{0}^{\infty}{e^{-st}f_{1,2}(n,t)}\,dt}$, we 
obtain
\begin{equation}
G_b(z) = \lim _{s\to 0}\Bigl(\mu_m F_1(z,s) + \mu_c F_2(z,s)\Bigr)\label{eq:Gb_single}
\end{equation}

\par

\begin{figure}[tb]
\resizebox{8cm}{!}{\includegraphics{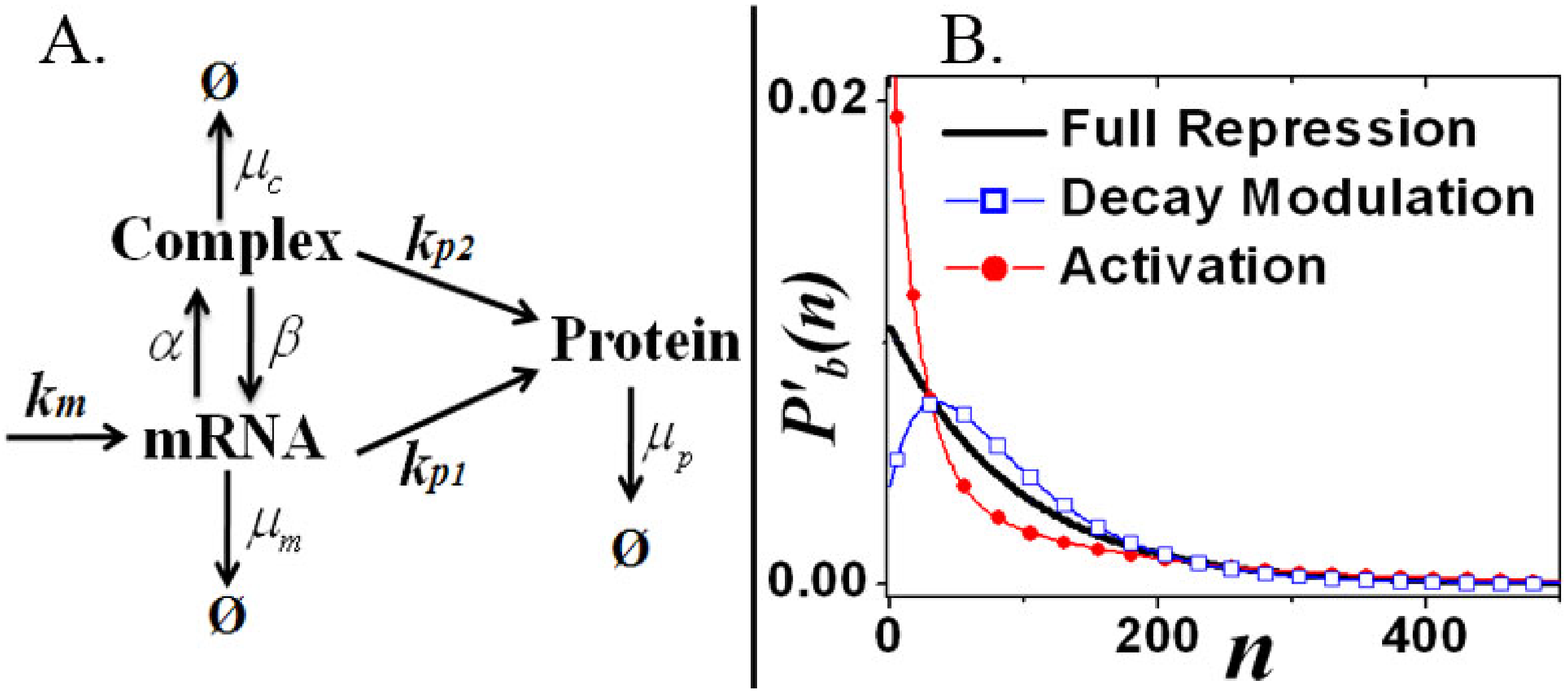}}
\caption{(A)Kinetic scheme for regulation of protein production by a
mRNA-binding regulator. (B)The protein burst distribution
$P^\prime_{b}(n)$ for full repression, decay modulation and activation
from Eq.(\ref{eq:Gz}).  In all three cases, the mean of the protein
burst distribution is kept the same.  The burst distribution for full
repression is identical to the geometric distribution with the same
mean, whereas $P^\prime_{b}(n)$ for decay modulation and activation
deviates significantly from the geometric distribution. Parameters for
decay modulation and activation are chosen such that
$\frac{\alpha}{\mu_m} = 5$, $\frac{\beta}{\mu_m} = 1$,
$\frac{\mu_c}{\mu_m} = 5$, $p_m =1$ and $\frac{\alpha}{\mu_m} =
\frac{\beta}{\mu_m} = 1$, $\frac{kp_2}{kp_1} = 4$, $p_m =1$
respectively.  \label{fig2}}
\end{figure}\noindent

We now consider the Telegraph scenario wherein multiple mRNAs can be
produced during a single burst. In this case, the probability of
having $m$ mRNAs in one burst, $P(m)$, is given by a geometric
distribution, conditional on the production of at least 1 mRNA
\cite{ingram08}
\begin{equation}
P(m)=(1-p_m)^{m-1}p_m. \label{eq:pm}
\end{equation}
Eq.(\ref{eq:pm}) serves as a general formula for the mRNA
burst distribution characterizing {\em both} Poisson and Telegraph
scenarios. The case $p_m=1$ correspond to a single mRNA produced every
burst (Poisson scenario); whereas if $p_m < 1$, the mean number of
mRNAs produced per burst $m_b(= 1/p_m)$ is greater than 1 (Telegraph
scenario). In general, each mRNA will produce a random number of
proteins drawn from the distribution $P_b(n)$ (the corresponding
generating function $G_b(z)$ is given by Eq.(\ref{eq:Gb_single})) and
furthermore the number of mRNAs in the burst is also a random variable
defined by the distribution Eq.(\ref{eq:pm}).  We denote the
distribution of proteins produced from {\em all} the mRNAs in the
burst by $P^\prime_{b}(n)$. The corresponding generating
function $G^\prime_{b}(z)$ is given by
\begin{equation}
G^\prime_{b}(z)=\frac{G_{b}(z)p_m}{1-G_{b}(z)(1-p_m)},\label{eq:Gz2}
\end{equation}

\par

Evaluation of the functions  $F_{1,2}(z,s)$ in combination with 
Eq.(\ref{eq:Gz2}) and Eq.(\ref{eq:Gb_single}) then leads to
the exact expression for $G^\prime_b(z)$, which can be written as

\begin{equation}
G^\prime_{b}(z)=X\frac{1-S_1}{z-S_1} + (1-X)\frac{1-S_2}{z-S_2} \nonumber
\end{equation}
where
\begin{eqnarray}
X &=&\frac{\sqrt{\Delta} -
\bigl(k_{p1}(\beta+\mu_c)+k_{p2}(\alpha- \mu_m p_m)\bigr)}{2\sqrt{\Delta}}
\nonumber \\
S_{1,2}&=&1+\frac{k_{p1}(\beta+ \mu_c)+k_{p2}(\alpha+\mu_m p_m)\pm\sqrt{\Delta}}{2k_{p1}k_{p2}}
\nonumber \\
\Delta &=&\bigl(k_{p1}(\beta+\mu_c)-k_{p2}(\alpha+\mu_m p_m)\bigr)^2\nonumber
\\ &+&4\alpha k_{p1}k_{p2}(\beta+\mu_c-\mu_cp_m). \label{eq:Gz}
\end{eqnarray}

The above expression indicates that the distribution of proteins
produced in a single burst ($P^\prime_{b}(n)$) can be expressed as a
weighted sum of two geometric distributions.
While the complete expression for $P^\prime_{b}(n)$ can thus be
derived from the results obtained, in some cases, the primary interest
is in derived quantities characterizing the noise in protein
distributions.  For example, several studies have focused on the noise
strength (or Fano Factor) $\sigma_b^2/n_b$ \cite{kaern05}.  For the protein
burst distribution $P^\prime_{b}(n)$, both the mean and the noise strength
can be obtained from the generating function as
\begin{eqnarray}
%n_b &=& \frac{\frac{k_{p1}}{p_m} + \frac{k_{p2}}{p_m} \frac{\alpha}{\mu_c + \beta}}{\mu_m + \frac{\mu_c \alpha}{\mu_c + \beta}} \nonumber \\
n_b &=& \frac{\frac{k_{p1}}{p_m} (\mu_c + \beta) + \frac{k_{p2}}{p_m} \alpha}{\mu_m (\mu_c + \beta) + \mu_c \alpha} \nonumber \\
\frac{\sigma_b^2}{n_b}&=&1+n_b\nonumber \\
&+&\frac{2 \alpha k_{p2}(k_{p2}\mu_m-k_{p1}\mu_c)}{(\alpha k_{p2}+k_{p1}\beta+k_{p1}\mu_c)(\alpha\mu_c+\beta\mu_m+\mu_c\mu_m)}\label{eq:fanofactor}
\end{eqnarray}
%The above expressions are in excellent agreement with results from 
%stochastic simulations using the Gillespie algorithm.
\par

It is noteworthy that Eq. (\ref{eq:Gz}) is valid for the most general choice 
of parameters. To gain additional insight, let us consider specific parameter 
choices of interest. For example, taking the limit $\alpha \to 0$ corresponds
to the unregulated protein burst distribution. In this case, we obtain
\begin {equation}
G^\prime_b(z) = \frac{\mu_m}{ \mu_m + \frac{k_{p1}}{p_m}(1-z)}, \label{eq:G_b_no_regulation}
\end {equation}
which corresponds to the generating function of a geometric
distribution in agreement with previous studies
\cite{yu06,friedman06,ingram08,swain09}. Of greater interest is the
effect of different modes of regulation.  While a generally accepted
model is that regulator binding prevents ribosome accesss
(i.e. $k_{p2}=0$), recent studies have shown that small RNAs can also
repress gene expression by binding in the coding region significantly
downstream of the ribosome binding site \cite{pfeiffer09}. In the
latter case, regulator binding is not expected to affect the
translation rate, but instead alters the mRNA decay rate. To explore
the effects of these different regulatory mechanisms on the noise in
protein distributions, we consider two special cases for the general
results derived above: 1) full repression ($k_{p2}=0$) and 2) decay
modulation ($k_{p2}= k_{p1}$, $u_{m} < u_{c}$).  \par

For full repression (in the limit $k_{p2}\to 0$), we have
\begin {equation}
G^\prime_b(z) = \frac{\mu_m + \frac{\mu_c \alpha}{\mu_c+\beta}}{\mu_m + \frac{\mu_c \alpha}{\mu_c+\beta} - \frac{k_{p1}}{p_m}(1-z)}\label{eq:repression}.
\end {equation}
The result is identical to Eq.(\ref{eq:G_b_no_regulation}) provided
that the mRNA degradation rate is rescaled from $\mu_m$ to $\mu_m +
\frac{\mu_c \alpha}{\mu_c+\beta}$. Thus the protein burst distribution
remains a geometric distribution but with a reduced mean due to lowering 
of the effective mRNA lifetime. This implies that regulation by full 
repression results in a protein burst distribution that is identical 
to that of an unregulated burst distribution with the same mean.
\par

On the other hand for regulation by decay modulation, the burst
distribution shows deviations from a geometric distribution (Fig. 2B).
To analyze this further,  let us focus 
on the noise strength $\sigma_b^2/n_b$ in Eq.(\ref{eq:fanofactor}) which, for 
decay modulation, is given by 
\begin{eqnarray}  
\frac{\sigma_b^2}{n_b}&=& 1+n_b+\frac{2 \alpha k_{p1}(1 - \theta_1)}{(\alpha + \beta + \theta_1 \mu_m)(\beta + \theta_1 (\alpha + \mu_m))}\nonumber \\
&=& 1+n_b + Q \label{eq:deviation}
\end{eqnarray}
where $\theta_1=\mu_c/\mu_m>1$ and the term $Q$ quantifies the
deviation from the geometric distribution ($Q=0$ for a geometric
distribution).  Thus for regulation by decay modulation, the noise
strength can be tuned by the parameter $\theta_1$ resulting in a burst
distribution with reduced variance when compared an unregulated burst
distribution with the same mean.  Eq. (\ref{eq:deviation}) indicates
that this reduction can be significant since the maximum magnitude for
$\frac{Q}{n_b}$ is 0.5.  Such a narrowing of the variance relative to
the mean has been previously proposed as a potential function for
small RNAs with important implications for canalization of gene
expression during development \cite{hornstein06}.

\par

The previous results for repression mechanisms can be contrasted with
the effect of post-transcriptional activation of gene expression.  The
burst distribution for activation also shows significant deviations
from a geometric distribution with the same mean (Fig. 2B).  For
activation due to increased protein production (with $\mu_c=\mu_m$),
the deviation $Q$ relative to the burst mean is given by:
\begin{equation}
\frac{Q}{n_b} = \frac{2 \alpha \theta_2 p_m \mu_m (\theta_2
-1)}{(\alpha \theta_2 + \beta +\mu_m)^2}.
\end{equation}
where $\theta_2 = kp_2/kp_1$. As $\theta_2 > 1$ for activation, the
noise will be greater than that of an unregulated burst distribution 
with the same mean. The value
of $\frac{Q}{n_b}$, depending on the choice of $\theta_1$ and
$\alpha$, can be made arbitraily large. Our results thus indicate that
activation of gene expression by small RNAs can potentially lead to
large variance in protein distribution, which in turn can give rise to
phenotypic heterogeneity that is often beneficial for the organism
\cite{fraser09}.  
\par

Eq.(\ref{eq:deviation}) also illustrates conditions under which the
Poisson and Telegraph scenarios can be distinguished based on
observations of protein burst distributions. The unregulated burst
distribution is geometric and thus completely determined by its mean
value $n_b = \frac{1}{p_m} (\frac{k_{p1}}{\mu_m})$.  Since there is
effectively one measurable quantity ($n_b$) for the burst
distribution, $p_m$ canot be determined given that
$\frac{k_{p1}}{\mu_m}$ is not known \cite{ingram08}. Hence the Poisson
and Telegraph scenarios cannot be distinguished in this case.
However, for the case of decay modulation (with $k_{p1}=k_{p2}$), we
have an additional measurable quantity: $Q$.  It is of interest to
note that $Q$ depends on $\frac{k_{p1}}{\mu_m}$, but is independent of
$p_m$. Thus, measurements of $Q$ and $n_b$ can  be used
to determine both $\frac{k_{p1}}{\mu_m}$ and $p_m$ and thereby to
discriminate between the Poisson and Telegraph scenarios.

\par

The argument above provides a means of determining $p_m$ provided the
interaction parameters such as $\alpha$, $\beta$ and $\theta_1$ are
known.  In general, these parameters are not known, however for
regulators such that the dissociation rate $\frac{\beta}{\mu_m} \to 0$, the
following protocol can be used to determine $p_m$: (i) Obtain
the mean protein burst levels without regulation, denoted by $n_0$.
(ii) Choose a certain regulator concentration. Obtain the mean protein
burst level $n_{b1}$ and the corresponding variance. Determine the
deviation from a geometric distribution as defined in
Eq.(\ref{eq:deviation}), which is denoted by $Q_1$ and let
$n_1=n_{b1}/n_0$. (iii) Change the concentration of the regulator,
which effectively changes the regulator binding rate $\alpha$. Repeat
step (ii) and obtain the corresponding quantities denoted by $Q_2$ and
$n_2=n_{b2}/n_0$.  Given the five quantities $n_{0,1,2}$ and
$Q_{1,2}$, the mean transcriptional burst size $m_b$($=1/p_m$) is given by:
\begin{eqnarray}
m_b &=& -2 n_0 n_1 n_2\frac{Q_1(1-n_2)-Q_2(1-n_1)}{Q_1 n_1 (1-n_2) -
Q_2 n_2 (1-n_1)} \nonumber \\ &\times&
\frac{(1-n_1)(1-n_2)(n_1-n_2)}{Q_1 n_1 (1-n_2)^2 - Q_2 n_2 (1-n_1)^2}
\label{eq:solve_m_b}
\end{eqnarray}
Using stochastic simulations, we have verified that the above
expression accurately predicts the degree of transcriptional bursting.
It should be noted that experimental approaches have been developed
recently for direct measurements of mRNA burst distributions
\cite{raj08a,raj09} and it would be informative to compare results
from these direct approaches with estimates from the above
protocol.

\par
%Specifically, we set up simulations based on the protocol described above 
%with an input mRNA burst distribution corresponding to Eq.(\ref{eq:pm})
%with a given mean $m_{b}$ (denoted by $m_b$ (Actual) in Fig. 3).
%The simulation output is the protein burst mean and variance
%for different values of the parameter $\alpha$. We then estimate the 
%mean mRNA burst size $m_b$ using Eq.(\ref{eq:solve_m_b}) which, as indicated 
%in Fig. 3, agrees well with the actual $m_b$ value.

Finally, we note that the result derived above lead to corresponding
analytical expressions for steady-state protein distributions over a
population of cells. Recent work has shown that, when protein
lifetimes are much longer than mRNA lifetimes, an effective Master
equation can be written down for proteins alone
\cite{friedman06,swain09}. In this approximation, given a geometric
distribution for protein bursts, the corresponding steady-state
protein distribution is a negative binomial distribution
\cite{swain09}.  Given the most general burst distribution obtained
above, we derive that the corresponding steady-state distribution is a
convolution of two negative binomial distributions. A detailed
analysis of the corresponding expressions for the mean and variance
will be presented elsewhere.

\par

In summary, we have derived analytical expressions which characterize
the noise in protein distributions for a stochastic model of
post-transcriptional regulation. It is noteworthy that the expressions
provide functional forms for the entire probability distribution (and
not just the mean and variance) for arbitrary parameter choices.  This
knowledge can be a useful input for approaches to infer cellular
mechanisms and parameters based on entire distributions
\cite{munsky09,warmflash08}. The results also provide insight into
how different mechanisms of post-transcriptional regulation can be
used to fine-tune the noise in stochastic gene expression
with potential implications for studies addressing on the evolutionary
importance of noise in biological systems \cite{cagatay09}.  In some
limits, the modulated burst distributions can be used to infer the degree of
transcriptional bursting and hence to determine the source of
intrinsic noise in gene expression.  The results derived can serve as
building blocks for future studies focusing on regulation of 
stochastic gene expression. \\

\begin{acknowledgments}
The authors acknowledge funding support from ICTAS, Virginia Tech. We thank Vlad Elgart and Andrew Fenley for helpful discussions.
\end{acknowledgments}

\bibliography{stochastic_modeling}

%Merlin.mbs v4.21 2009-07-09.
\begin{thebibliography}{10}%
\makeatletter
\providecommand \@ifxundefined [1]{%
 \ifx #1\undefined \expandafter \@firstoftwo
 \else \expandafter \@secondoftwo
\fi
}%
\providecommand \@ifnum [1]{%
 \ifnum #1\expandafter \@firstoftwo
 \else \expandafter \@secondoftwo
\fi
}%
\providecommand \enquote [1]{``#1''}%
\providecommand \bibnamefont  [1]{#1}%
\providecommand \bibfnamefont [1]{#1}%
\providecommand \citenamefont [1]{#1}%
\providecommand\href[0]{\@sanitize\@href}%
\providecommand\@href[1]{\endgroup\@@startlink{#1}\endgroup\@@href}%
\providecommand\@@href[1]{#1\@@endlink}%
\providecommand \@sanitize [0]{\begingroup\catcode`\&12\catcode`\#12\relax}%
\@ifxundefined \pdfoutput {\@firstoftwo}{%
 \@ifnum{\z@=\pdfoutput}{\@firstoftwo}{\@secondoftwo}%
}{%
 \providecommand\@@startlink[1]{\leavevmode\special{html:<a href="#1">}}%
 \providecommand\@@endlink[0]{\special{html:</a>}}%
}{%
 \providecommand\@@startlink[1]{%
  \leavevmode
  \pdfstartlink
   attr{/Border[0 0 1 ]/H/I/C[0 1 1]}%
   user{/Subtype/Link/A<</Type/Action/S/URI/URI(#1)>>}%
  \relax
 }%
 \providecommand\@@endlink[0]{\pdfendlink}%
}%
\providecommand \url  [0]{\begingroup\@sanitize \@url }%
\providecommand \@url [1]{\endgroup\@href {#1}{\urlprefix}}%
\providecommand \urlprefix [0]{URL }%
\providecommand \Eprint[0]{\href }%
\@ifxundefined \urlstyle {%
  \providecommand \doi [1]{doi:\discretionary{}{}{}#1}%
}{%
  \providecommand \doi [0]{doi:\discretionary{}{}{}\begingroup
  \urlstyle{rm}\Url }%
}%
\providecommand \doibase [0]{http://dx.doi.org/}%
\providecommand \Doi[1]{\href{\doibase#1}}%
\providecommand \bibAnnote [3]{%
  \BibitemShut{#1}%
  \begin{quotation}\noindent
    \textsc{Key:}\ #2\\\textsc{Annotation:}\ #3%
  \end{quotation}%
}%
\providecommand \bibAnnoteFile [2]{%
  \IfFileExists{#2}{\bibAnnote {#1} {#2} {\input{#2}}}{}%
}%
\providecommand \typeout [0]{\immediate \write \m@ne }%
\providecommand \selectlanguage [0]{\@gobble}%
\providecommand \bibinfo [0]{\@secondoftwo}%
\providecommand \bibfield [0]{\@secondoftwo}%
\providecommand \translation [1]{[#1]}%
\providecommand \BibitemOpen[0]{}%
\providecommand \bibitemStop [0]{}%
\providecommand \bibitemNoStop [0]{.\EOS\space}%
\providecommand \EOS [0]{\spacefactor3000\relax}%
\providecommand \BibitemShut [1]{\csname bibitem#1\endcsname}%
%</preamble>
\bibitem{kaern05}%
  \BibitemOpen
  \bibfield{author}{%
  \bibinfo {author} {\bibfnamefont{M.}~\bibnamefont{Kaern}}, \bibinfo {author}
  {\bibfnamefont{T.~C.}\ \bibnamefont{Elston}}, \bibinfo {author}
  {\bibfnamefont{W.~J.}\ \bibnamefont{Blake}},\ and\ \bibinfo {author}
  {\bibfnamefont{J.~J.}\ \bibnamefont{Collins}},\ }%
  \bibfield{journal}{%
  \bibinfo {journal} {Nat Rev Genet}\ }%
  \textbf{\bibinfo {volume} {6}},\ \bibinfo {pages} {451} (\bibinfo {year}
  {2005})%
  \bibAnnoteFile{NoStop}{kaern05}%
\bibitem{waters09}%
  \BibitemOpen
  \bibfield{author}{%
  \bibinfo {author} {\bibfnamefont{L.}~\bibnamefont{Waters}}\ and\ \bibinfo
  {author} {\bibfnamefont{G.}~\bibnamefont{Storz}},\ }%
  \bibfield{journal}{%
  \bibinfo {journal} {{Cell}}\ }%
  \textbf{\bibinfo {volume} {{136}}},\ \bibinfo {pages} {615} (\bibinfo {year}
  {{2009}})%
  \bibAnnoteFile{NoStop}{waters09}%
\bibitem{kaufmann07}%
  \BibitemOpen
  \bibfield{author}{%
  \bibinfo {author} {\bibfnamefont{B.~B.}\ \bibnamefont{Kaufmann}}\ and\
  \bibinfo {author} {\bibfnamefont{A.}~\bibnamefont{van Oudenaarden}},\ }%
  \bibfield{journal}{%
  \bibinfo {journal} {Curr Opin Genet Dev}\ }%
  \textbf{\bibinfo {volume} {17}},\ \bibinfo {pages} {107} (\bibinfo {year}
  {2007})%
  \bibAnnoteFile{NoStop}{kaufmann07}%
\bibitem{thattai01}%
  \BibitemOpen
  \bibfield{author}{%
  \bibinfo {author} {\bibfnamefont{M.}~\bibnamefont{Thattai}}\ and\ \bibinfo
  {author} {\bibfnamefont{A.}~\bibnamefont{van Oudenaarden}},\ }%
  \bibfield{journal}{%
  \bibinfo {journal} {Proc Natl Acad Sci U S A}\ }%
  \textbf{\bibinfo {volume} {98}},\ \bibinfo {pages} {8614} (\bibinfo {year}
  {2001})%
  \bibAnnoteFile{NoStop}{thattai01}%
\bibitem{raser04}%
  \BibitemOpen
  \bibfield{author}{%
  \bibinfo {author} {\bibfnamefont{J.}~\bibnamefont{Raser}}\ and\ \bibinfo
  {author} {\bibfnamefont{E.}~\bibnamefont{O'Shea}},\ }%
  \bibfield{journal}{%
  \bibinfo {journal} {{Science}}\ }%
  \textbf{\bibinfo {volume} {{304}}},\ \bibinfo {pages} {1811} (\bibinfo {year}
  {{2004}})%
  \bibAnnoteFile{NoStop}{raser04}%
\bibitem{karmakar04}%
  \BibitemOpen
  \bibfield{author}{%
  \bibinfo {author} {\bibfnamefont{R.}~\bibnamefont{Karmakar}}\ and\ \bibinfo
  {author} {\bibfnamefont{I.}~\bibnamefont{Bose}},\ }%
  \bibfield{journal}{%
  \bibinfo {journal} {{Phys. Biol.}}\ }%
  \textbf{\bibinfo {volume} {{1}}},\ \bibinfo {pages} {197} (\bibinfo {year}
  {{2004}})%
  \bibAnnoteFile{NoStop}{karmakar04}%
\bibitem{paulsson05}%
  \BibitemOpen
  \bibfield{author}{%
  \bibinfo {author} {\bibfnamefont{J.}~\bibnamefont{Paulsson}},\ }%
  \bibfield{journal}{%
  \bibinfo {journal} {{Phys Of Life Rev}}\ }%
  \textbf{\bibinfo {volume} {{2}}},\ \bibinfo {pages} {157} (\bibinfo {year}
  {{2005}})%
  \bibAnnoteFile{NoStop}{paulsson05}%
\bibitem{iyer09}%
  \BibitemOpen
  \bibfield{author}{%
  \bibinfo {author} {\bibfnamefont{S.}~\bibnamefont{Iyer-Biswas}}, \bibinfo
  {author} {\bibfnamefont{F.}~\bibnamefont{Hayot}},\ and\ \bibinfo {author}
  {\bibfnamefont{C.}~\bibnamefont{Jayaprakash}},\ }%
  \bibfield{journal}{%
  \bibinfo {journal} {Phys. Rev. E}\ }%
  \textbf{\bibinfo {volume} {79}},\ \bibinfo {pages} {031911} (\bibinfo {year}
  {2009})%
  \bibAnnoteFile{NoStop}{iyer09}%
\bibitem{bareven06}%
  \BibitemOpen
  \bibfield{author}{%
  \bibinfo {author} {\bibfnamefont{A.}~\bibnamefont{Bar-Even}}, \bibinfo
  {author} {\bibfnamefont{J.}~\bibnamefont{Paulsson}}, \bibinfo {author}
  {\bibfnamefont{N.}~\bibnamefont{Maheshri}}, \bibinfo {author}
  {\bibfnamefont{M.}~\bibnamefont{Carmi}}, \bibinfo {author}
  {\bibfnamefont{E.}~\bibnamefont{O'Shea}}, \bibinfo {author}
  {\bibfnamefont{Y.}~\bibnamefont{Pilpel}},\ and\ \bibinfo {author}
  {\bibfnamefont{N.}~\bibnamefont{Barkai}},\ }%
  \bibfield{journal}{%
  \bibinfo {journal} {Nat Genet}\ }%
  \textbf{\bibinfo {volume} {38}},\ \bibinfo {pages} {636} (\bibinfo {year}
  {2006})%
  \bibAnnoteFile{NoStop}{bareven06}%
\bibitem{newman06}%
  \BibitemOpen
  \bibfield{author}{%
  \bibinfo {author} {\bibfnamefont{J.~R.~S.}\ \bibnamefont{Newman}}, \bibinfo
  {author} {\bibfnamefont{S.}~\bibnamefont{Ghaemmaghami}}, \bibinfo {author}
  {\bibfnamefont{J.}~\bibnamefont{Ihmels}}, \bibinfo {author}
  {\bibfnamefont{D.~K.}\ \bibnamefont{Breslow}}, \bibinfo {author}
  {\bibfnamefont{M.}~\bibnamefont{Noble}}, \bibinfo {author}
  {\bibfnamefont{J.~L.}\ \bibnamefont{DeRisi}},\ and\ \bibinfo {author}
  {\bibfnamefont{J.~S.}\ \bibnamefont{Weissman}},\ }%
  \bibfield{journal}{%
  \bibinfo {journal} {{Nature}}\ }%
  \textbf{\bibinfo {volume} {{441}}},\ \bibinfo {pages} {840} (\bibinfo {year}
  {{2006}})%
  \bibAnnoteFile{NoStop}{newman06}%
\bibitem{yu06}%
  \BibitemOpen
  \bibfield{author}{%
  \bibinfo {author} {\bibfnamefont{J.}~\bibnamefont{Yu}}, \bibinfo {author}
  {\bibfnamefont{J.}~\bibnamefont{Xiao}}, \bibinfo {author}
  {\bibfnamefont{X.}~\bibnamefont{Ren}}, \bibinfo {author}
  {\bibfnamefont{K.}~\bibnamefont{Lao}},\ and\ \bibinfo {author}
  {\bibfnamefont{X.~S.}\ \bibnamefont{Xie}},\ }%
  \bibfield{journal}{%
  \bibinfo {journal} {Science}\ }%
  \textbf{\bibinfo {volume} {311}},\ \bibinfo {pages} {1600} (\bibinfo {year}
  {2006})%
  \bibAnnoteFile{NoStop}{yu06}%
\bibitem{ingram08}%
  \BibitemOpen
  \bibfield{author}{%
  \bibinfo {author} {\bibfnamefont{P.~J.}\ \bibnamefont{Ingram}}, \bibinfo
  {author} {\bibfnamefont{M.~P.~H.}\ \bibnamefont{Stumpf}},\ and\ \bibinfo
  {author} {\bibfnamefont{J.}~\bibnamefont{Stark}},\ }%
  \bibfield{journal}{%
  \bibinfo {journal} {{PLoS Comp Biol}}\ }%
  \textbf{\bibinfo {volume} {{4}}} (\bibinfo {year} {{2008}})%
  \bibAnnoteFile{NoStop}{ingram08}%
\bibitem{levine07}%
  \BibitemOpen
  \bibfield{author}{%
  \bibinfo {author} {\bibfnamefont{E.}~\bibnamefont{Levine}}, \bibinfo {author}
  {\bibfnamefont{Z.}~\bibnamefont{Zhang}}, \bibinfo {author}
  {\bibfnamefont{T.}~\bibnamefont{Kuhlman}},\ and\ \bibinfo {author}
  {\bibfnamefont{T.}~\bibnamefont{Hwa}},\ }%
  \bibfield{journal}{%
  \bibinfo {journal} {PLoS Biol}\ }%
  \textbf{\bibinfo {volume} {5}},\ \bibinfo {pages} {e229} (\bibinfo {year}
  {2007})%
  \bibAnnoteFile{NoStop}{levine07}%
\bibitem{mitarai07}%
  \BibitemOpen
  \bibfield{author}{%
  \bibinfo {author} {\bibfnamefont{N.}~\bibnamefont{Mitarai}}, \bibinfo
  {author} {\bibfnamefont{A.~M.}\ \bibnamefont{Andersson}}, \bibinfo {author}
  {\bibfnamefont{S.}~\bibnamefont{Krishna}}, \bibinfo {author}
  {\bibfnamefont{S.}~\bibnamefont{Semsey}},\ and\ \bibinfo {author}
  {\bibfnamefont{K.}~\bibnamefont{Sneppen}},\ }%
  \bibfield{journal}{%
  \bibinfo {journal} {Phys Biol}\ }%
  \textbf{\bibinfo {volume} {4}},\ \bibinfo {pages} {164} (\bibinfo {year}
  {2007})%
  \bibAnnoteFile{NoStop}{mitarai07}%
\bibitem{mehta08}%
  \BibitemOpen
  \bibfield{author}{%
  \bibinfo {author} {\bibfnamefont{P.}~\bibnamefont{Mehta}}, \bibinfo {author}
  {\bibfnamefont{S.}~\bibnamefont{Goyal}},\ and\ \bibinfo {author}
  {\bibfnamefont{N.~S.}\ \bibnamefont{Wingreen}},\ }%
  \bibfield{journal}{%
  \bibinfo {journal} {{Mol Sys Biol}}\ }%
  \textbf{\bibinfo {volume} {{4}}} (\bibinfo {year} {{2008}})%
  \bibAnnoteFile{NoStop}{mehta08}%
\bibitem{berg78}%
  \BibitemOpen
  \bibfield{author}{%
  \bibinfo {author} {\bibfnamefont{O.~G.}\ \bibnamefont{Berg}},\ }%
  \bibfield{journal}{%
  \bibinfo {journal} {J Theor Biol}\ }%
  \textbf{\bibinfo {volume} {71}},\ \bibinfo {pages} {587} (\bibinfo {year}
  {1978})%
  \bibAnnoteFile{NoStop}{berg78}%
\bibitem{friedman06}%
  \BibitemOpen
  \bibfield{author}{%
  \bibinfo {author} {\bibfnamefont{N.}~\bibnamefont{Friedman}}, \bibinfo
  {author} {\bibfnamefont{L.}~\bibnamefont{Cai}},\ and\ \bibinfo {author}
  {\bibfnamefont{X.~S.}\ \bibnamefont{Xie}},\ }%
  \bibfield{journal}{%
  \bibinfo {journal} {Phys Rev Lett}\ }%
  \textbf{\bibinfo {volume} {97}},\ \bibinfo {pages} {168302} (\bibinfo {year}
  {2006})%
  \bibAnnoteFile{NoStop}{friedman06}%
\bibitem{swain09}%
  \BibitemOpen
  \bibfield{author}{%
  \bibinfo {author} {\bibfnamefont{V.}~\bibnamefont{Shahrezaei}}\ and\ \bibinfo
  {author} {\bibfnamefont{P.~S.}\ \bibnamefont{Swain}},\ }%
  \bibfield{journal}{%
  \bibinfo {journal} {{Proc Natl Acad Sci USA}}\ }%
  \textbf{\bibinfo {volume} {{105}}},\ \bibinfo {pages} {17256} (\bibinfo
  {month} {{NOV 11}}\ \bibinfo {year} {{2008}})%
  \bibAnnoteFile{NoStop}{swain09}%
\bibitem{pfeiffer09}%
  \BibitemOpen
  \bibfield{author}{%
  \bibinfo {author} {\bibfnamefont{V.}~\bibnamefont{Pfeiffer}}, \bibinfo
  {author} {\bibfnamefont{K.}~\bibnamefont{Papenfort}}, \bibinfo {author}
  {\bibfnamefont{S.}~\bibnamefont{Lucchini}}, \bibinfo {author}
  {\bibfnamefont{J.~C.~D.}\ \bibnamefont{Hinton}},\ and\ \bibinfo {author}
  {\bibfnamefont{J.}~\bibnamefont{Vogel}},\ }%
  \bibfield{journal}{%
  \bibinfo {journal} {{Nat. Struct. Mol. Biol.}}\ }%
  \textbf{\bibinfo {volume} {{16}}},\ \bibinfo {pages} {840} (\bibinfo {month}
  {{AUG}}\ \bibinfo {year} {{2009}})%
  \bibAnnoteFile{NoStop}{pfeiffer09}%
\bibitem{hornstein06}%
  \BibitemOpen
  \bibfield{author}{%
  \bibinfo {author} {\bibfnamefont{E.}~\bibnamefont{Hornstein}}\ and\ \bibinfo
  {author} {\bibfnamefont{N.}~\bibnamefont{Shomron}},\ }%
  \bibfield{journal}{%
  \bibinfo {journal} {{Nat. Genet.}}\ }%
  \textbf{\bibinfo {volume} {{38}}},\ \bibinfo {pages} {S20} (\bibinfo {year}
  {{2006}})%
  \bibAnnoteFile{NoStop}{hornstein06}%
\bibitem{fraser09}%
  \BibitemOpen
  \bibfield{author}{%
  \bibinfo {author} {\bibfnamefont{D.}~\bibnamefont{Fraser}}\ and\ \bibinfo
  {author} {\bibfnamefont{M.}~\bibnamefont{Kaern}},\ }%
  \bibfield{journal}{%
  \bibinfo {journal} {{Mol. Microb.}}\ }%
  \textbf{\bibinfo {volume} {{71}}},\ \bibinfo {pages} {1333} (\bibinfo {year}
  {{2009}})%
  \bibAnnoteFile{NoStop}{fraser09}%
\bibitem{raj08a}%
  \BibitemOpen
  \bibfield{author}{%
  \bibinfo {author} {\bibfnamefont{A.}~\bibnamefont{Raj}}, \bibinfo {author}
  {\bibfnamefont{P.}~\bibnamefont{van~den Bogaard}}, \bibinfo {author}
  {\bibfnamefont{S.~A.}\ \bibnamefont{Rifkin}}, \bibinfo {author}
  {\bibfnamefont{A.}~\bibnamefont{van Oudenaarden}},\ and\ \bibinfo {author}
  {\bibfnamefont{S.}~\bibnamefont{Tyagi}},\ }%
  \bibfield{journal}{%
  \bibinfo {journal} {{Nat. Meth. }}\ }%
  \textbf{\bibinfo {volume} {{5}}},\ \bibinfo {pages} {877} (\bibinfo {year}
  {{2008}})%
  \bibAnnoteFile{NoStop}{raj08a}%
\bibitem{raj09}%
  \BibitemOpen
  \bibfield{author}{%
  \bibinfo {author} {\bibfnamefont{A.}~\bibnamefont{Raj}}\ and\ \bibinfo
  {author} {\bibfnamefont{A.}~\bibnamefont{van Oudenaarden}},\ }%
  \bibfield{journal}{%
  \bibinfo {journal} {{Ann. Rev. Biophys.}}\ }%
  \textbf{\bibinfo {volume} {{38}}},\ \bibinfo {pages} {255} (\bibinfo {year}
  {{2009}})%
  \bibAnnoteFile{NoStop}{raj09}%
\bibitem{munsky09}%
  \BibitemOpen
  \bibfield{author}{%
  \bibinfo {author} {\bibfnamefont{B.}~\bibnamefont{Munsky}}, \bibinfo {author}
  {\bibfnamefont{B.}~\bibnamefont{Trinh}},\ and\ \bibinfo {author}
  {\bibfnamefont{M.}~\bibnamefont{Khammash}},\ }%
  \bibfield{journal}{%
  \bibinfo {journal} {{Mol. Sys. Biol.}}\ }%
  \textbf{\bibinfo {volume} {{5}}} (\bibinfo {year} {{2009}})%
  \bibAnnoteFile{NoStop}{munsky09}%
\bibitem{warmflash08}%
  \BibitemOpen
  \bibfield{author}{%
  \bibinfo {author} {\bibfnamefont{A.}~\bibnamefont{Warmflash}}\ and\ \bibinfo
  {author} {\bibfnamefont{A.~R.}\ \bibnamefont{Dinner}},\ }%
  \bibfield{journal}{%
  \bibinfo {journal} {{Proc. Nat. Acad. Sci. USA}}\ }%
  \textbf{\bibinfo {volume} {{105}}},\ \bibinfo {pages} {17262} (\bibinfo
  {year} {{2008}})%
  \bibAnnoteFile{NoStop}{warmflash08}%
\bibitem{cagatay09}%
  \BibitemOpen
  \bibfield{author}{%
  \bibinfo {author} {\bibfnamefont{T.}~\bibnamefont{Cagatay}}, \bibinfo
  {author} {\bibfnamefont{M.}~\bibnamefont{Turcotte}}, \bibinfo {author}
  {\bibfnamefont{M.~B.}\ \bibnamefont{Elowitz}}, \bibinfo {author}
  {\bibfnamefont{J.}~\bibnamefont{Garcia-Ojalvo}},\ and\ \bibinfo {author}
  {\bibfnamefont{G.~M.}\ \bibnamefont{Suel}},\ }%
  \bibfield{journal}{%
  \bibinfo {journal} {{Cell}}\ }%
  \textbf{\bibinfo {volume} {{139}}},\ \bibinfo {pages} {512} (\bibinfo {year}
  {{2009}})%
  \bibAnnoteFile{NoStop}{cagatay09}%
\end{thebibliography}%

\end{document}